# Text Classification using Association Rule with a Hybrid Concept of Naive Bayes Classifier and Genetic Algorithm

S. M. Kamruzzaman, Farhana Haider and Ahmed Ryadh Hasan[†]

Department of Computer Science and Engineering

International Islamic University Chittagong, Chittagong-4203, Bangladesh

Email: smk_iiuc@yahoo.com, farhanahdr@yahoo.com

† School of Communication, Independent University Bangladesh, Email: ryadh78@yahoo.com

**Abstract**

*Text classification is the automated assignment of natural language texts to predefined categories based on their content. Text classification is the primary requirement of text retrieval systems, which retrieve texts in response to a user query, and text understanding systems, which transform text in some way such as producing summaries, answering questions or extracting data. Now a day the demand of text classification is increasing tremendously. Keeping this demand into consideration, new and updated techniques are being developed for the purpose of automated text classification. This paper presents a new algorithm for text classification. Instead of using words, word relation i.e association rules is used to derive feature set from pre-classified text documents. The concept of Naive Bayes Classifier is then used on derived features and finally a concept of Genetic Algorithm has been added for final classification. A system based on the proposed algorithm has been implemented and tested. The experimental results show that the proposed system works as a successful text classifier.*

## 1. INTRODUCTION

It has been established that classification is the core requirement of any information retrieval system. A text classifier retrieves texts to sense a user query, which in other sense retrieves the summary or domain of the raw text [2]. With the existing algorithms, a number of newly established processes are involving in the automation of text classification [4] [12] [13]. It has been observed that for the purpose of text classification the concept of association rule is very well known. Association rule mining [1] finds interesting association or relationships among a large set of data items [5]. These relationships can help in many decision making process. On the other hand, the Naive Bayes classifier uses the maximum a posteriori estimation for learning a classifier. It assumes that the occurrence of each word in a document is conditionally independent of all other words in that document given its class [6] and Genetic Algorithm attempts to incorporate ideas of natural evolution. It starts with an initial population which is created consisting of randomly generated rules. Each rule can be represented by a string of bits. Based on the notion of survival of the fittest, a new population is formed to consist of the fittest rules in the current population, as well as offspring of these rules. Typically, the fitness of a rule is assessed by its classification accuracy on a set of training examples.

## 2. PREVIOUS WORK

Researchers showed a number of techniques for text classification. In this paper we have considered the most recently established efficient approaches.

### 2.1 Works on Association Rule

#### 2.1.1 With Naive Bayes Classifier

Research on Text Classification using the Concept of Association Rule of Data Mining where Naive Bayes Classifier was used to classify text finally showed the dependability of the Naive Bayes Classifier with association rules [3] [5]. But since this method ignores the concept of calculation of negative example for any specific class determination, the accuracy may fall in some cases. For classifying a text it just calculates the probability of different class with the probability values of the matched set while ignoring the unmatched sets. As a result, if test set matches with a rule set which weak probability to the actual class may cause wrong classification.

#### 2.1.2 With Decision Tree

Text classification using decision tree showed an acceptable accuracy using 76% training data of total data set [11], while it is possible to achieve good accuracy using only 40 to 50% of total data sets [9].

### 2.2 Works on Genetic Algorithm

Text Classification based on Genetic Algorithm showed satisfactory performance using 69% training data while this process requires the time consuming steps to classify the texts [6]. Here, positive and negative files are used for each class and final weight is achieved by taking the result of subtraction from relative positive weight to relative negative weight. Each class requires a long calculation steps [4][11].

## 3. DETAILED VIEW

## 3.1 Data Mining

Simply stated, data mining refers to extracting or "mining" knowledge from large amounts data. Data mining is also treated as Knowledge Discovery in Databases, or KDD. In fact, data mining is the automated extraction of patterns representing knowledge implicitly stored in the large databases, data warehouses and other massive information stores [7]. Standard data mining methods may be integrated with information retrieval techniques and the construction or use of hierarchies specifically for text data. Text databases are databases that contain word descriptions for objects. These word descriptions are usually not simple keywords but rather long sentences or paragraphs, such as product specifications, error or bug reports, warning messages, summary reports, notes, or other documents. Data mining can do extraction of required information from text database. For this purpose the raw sentences or paragraphs are needed to be deal efficiently, since raw text data does not represent keywords directly. Hence, for effective evaluation, proper mining of raw data is needed. The widely used and well-known data mining functionalities are characterization and discrimination, content-based analysis, association analysis, classification and prediction [7], cluster analysis, outlier analysis, evolution analysis. For our text classification purpose we have used association analysis for generating associative word sets.

## 3.2 Association Rule

In short association rule is based on associated relationships. An assumption can give a mathematical view on association rule. Let $J = \{i1, i2, i3,...,im\}$ be a set of items. Let D, the task-relevant data, be a set of database transactions where each transaction T is a set of items such that T⊆J. Each transaction is associated with an identifier, called TID. Let A be a set of items. A transaction T is said to contain A if and only if A⊆T. An association rule is an implication of the form A⇒B, where A ⊂ J, B ⊂ J and A∩B = Æ. The rule A⇒B holds in the transaction set D with support s, where s is the percentage of transactions in D that contain A B (i.e. both A and B). This is taken to be the probability, P(A B). The rule A⇒B has confidence c in the transaction set D if c is the percentage of transactions in D containing A that also contain B. This is taken to be the conditional probability. P (B|A). That is, support (A⇒B) = P (A B), confidence (A⇒B) = P (B|A) = [support count(A B)/support count(A)].

Rules that satisfy both a minimum support threshold (min_sup) and a minimum confidence threshold (min_conf) are called strong. A set of items is referred to as an itemset. An itemset that contains k items is a k-itemset. The occurrence frequency of an itemset is the number of transactions that contain the itemse. An itemset satisfies minimum support if the occurrence frequency of the itemset is greater than or equal to the product of min_sup and the total number of transactions in D. The number of transaction required for the item set to satisfy minimum support is therefore referred to as the minimum support count. If an itemset satisfies minimum support then it is a frequent itemset [2] [7].

Association Mining is a two-step process.

1. Find all frequent itemsets: By definition, each of these itemsets will occur at least as frequently as predefined minimum support count.

2. Generating strong Association Rules from the frequent itemsets: By definition, these rules must satisfy minimum confidence.

## 3.3 The Apriori Algorithm

Apriori [1] is an influential algorithm for mining frequent itemsets. The algorithm is named such since it uses prior knowledge of frequent itemset properties.

Apriory employs an iterative approach known as a level-wise search, where k-itemsets are used to explore (k+1)-itemsets. First, the set of frequent 1-itemsets is found. This set is denoted L1. L1 is used to find L2, the set of frequent 2-itemsets, which is used to find L3, and so on, until no more frequent k-itemsets can be found. The finding of each Lk requires one full scan of the database. An important property called Apriori property, based on the observation is that, if an itemset I is not frequent, that is, P (I) < min_sup then if an item A is added to the itemset I, the resulting itemset (i.e., I A) cannot occur more frequently than I. Therefore, I A is not frquent either, that is, P (I A) < min_sup.

**The Apriori algorithm acts on two steps:**

**The Join Step:**

To find Lk, a set of candidate k-itemsets is generated by joining Lk-1 with itself. This set of candidates is denoted by Ck. Let l1 and l2 be itemsets in Lk-1, then l1 and l2 are joinable if their first (k-2) items are in common.

**The Prune Step:**

Ck is the superset of Lk. A scan of the database to determine the count of each candidate in Ck would result in the determination of Lk (itemsets having a count no less than minimum support in Ck). But this scan and computation can be reduced by applying the Apriory property. Any (k-1) itemsets that is not frequent cannot be a subset of frequent k itemset. Hence if any (k-1) subset of a candidate k-itemset is not in Lk-1, then the candidate cannot be frequent either and so can be removed from Ck.

## 3.4 Naive Bayes Classifier

In applying Naive Bayes classifier, each word position in a document is defined as an attribute and the value of that attribute to be the english word found in that position. Naive Bayes classification is given by:

$$V_{NB} = \text{argmax } P(V_j)\prod P(a_j | V_j)$$

To summarize, the Naive Bayes classification VNB is the classification that maximizes the probability of observing the words that were actually found in the example documents, subject to the usual Naive Bayes independence assumption. The first term can be estimated based on the fraction of each class in the training data. For estimating the second term the following equation is used:

$$\frac{n_k + 1}{n + |vocabulary|} \quad \quad \quad (1)$$

Where n is the total number of word positions in all training examples whose target value is Vj, nk is the number of items that word is found among these n word positions, and |vocabulary| is the total number of distinct words found within training data.

### 3.5 Genetic Algorithm in Text Classification

In general, genetic algorithm starts with an initial population consisting of randomly generated rules. Each rule can be represented by a string of bits. For example, suppose the samples in a given training set are described by two Boolean attributes, A1 and A2, and that there are two classes, C1 and C2. The rule "IF A1 AND NOT A2 THEN C2" can be encoded as the bit string "100", where the two leftmost bits represent attributes A1 and A2, respectively and the rightmost bit represents the class. Similarly, the rule "IF NOT A1 AND NOT A2 THEN C1" can be encoded as the bit string "001", If an attribute has k-values, where k>2, then k-bits may be used to encode the attribute's values. Classes can be encoded in a similar fashion.

Based on the notion of survival of the fittest, a new population is formed to consist of the fittest rules in the current population, as well as offspring of these rules. Typically, the fitness of a rule is assessed by its classification accuracy on a set of training examples. Applying genetic operators such as crossover and mutation creates offspring. In crossover, sub strings from pairs of the rules are swapped to form new pairs of rules. In mutation, randomly selected bits in a rule's string are inverted. The process of generating new populations based on prior populations of rules continues until a population P "evolves" where each rule in P satisfies a pre-specified fitness threshold.

In our experiment, we have considered three classes of text: Educational Engineering (EDE), Algorithm (ALG) and Artificial Intelligence (AI). For generating rules of a particular group we give positive and negative examples for it. For example, when we intend to learn the rules for classifying text for EDE, all examples of EDE act as positive examples and the rest from other groups act as negative examples.

## 4. THE PROPOSED METHOD
### 4.1 Preparing Text for Classification

Abstracts from different research papers are considered as training document for developing a model for classifying new documents of unknown class. Some of the abstracts are collected from the proceedings of ICCIT 2002 and the rest of them are from World Wide Web. Three classes of papers from Educational Engineering (EDE), Algorithm (ALG) and Artificial Intelligence (AI) are considered as training documents. Total 103 abstracts (27 from ALG, 14 from EDE and 62 from AI) have been used for the experiment. Each abstract is considered as a transaction in the text data. So number of abstracts is equal to the number of transactions in the transaction set (text data). The next step is to clean the text data by removing unnecessary words. To clean the text we have considered only the keywords. As we know, highly frequent words, such as determiners and prepositions, are not considered to be content words because they appear in virtually every document [10]. Unlike considering all words in a text we have considered only those words that are related to the subject of the text. A filtering process is adopted in order to remove unnecessary words. For this keyword extraction process we dropped the common unnecessary words like am, is, are, to, from...etc. and also dropped all kinds of punctuations and stop words. Singular and plural form of a word is considered same and keeping the singular form in the text. Finally, the remaining frequent words are considered as a single transaction data in the set of database transaction. This process is applied to all text data before applying association mining to the transaction database.

**Let an abstract:**

*Let G = (V, E) be a connected graph, and let X be a vertex subset of G. Let f be a mapping from X to the set of natural numbers such that f(x) ³ 2 for all x in X. A degree restricted spanning tree is a spanning tree T of G such that f(x) £ degT(x) for all x belongs X, where degT(x) denotes the degree of a vertex x in T. In this paper, we show that the decision problem "whether there exists a degree restricted spanning tree in G" is NP-complete. We also give a restricted proof of a conjecture, provided by Kaneko and Yoshimoto, on the existence of such a spanning tree in general graphs. Finally, we present a polynomial-time algorithm to find a degree restricted spanning tree of a graph satisfying the conditions presented in the restricted proof of the conjecture.*

Keywords extracted from this abstract are: Spanning, tree, bipartite, graph

This keyword extraction process is applied to all the abstracts.

### 4.2 Deriving Associated Word Sets

Considering each keyword set i.e. each abstract as a transaction, we generated a list of maximum length sets applying the Apriori algorithm. The support and confidence is set to 0.05 and 0.75 respectively. A partial list of the generated large word set with their occurrence frequency is illustrated in the following table.

**Table 1:** Word set with occurrence frequency

| Maximum Length Set | No of Occurrence | | |
|---|---|---|---|
| | EDE | ALG | AI |
| neural, network | | | 5 |
| gray, code | | 4 | |
| set, length | | 2 | |
| expectation, teacher, found, significant | 2 | | |
| education, level, test, significant, difference | 2 | | |
| significant, study, result | 3 | | |
| total, score, result | 2 | | |
| student, achievement, data, revealed | 2 | | |
| ratio, hole | 1 | | |
| gain, dependencies | | | 2 |
| lossless, compression | | | 2 |

### 4.3 Setting Associated Word Set With Probability Value

From the generated word set after applying association mining on training data we have found the following information: total number of word set is 27, total number of word set from Educational Engineering (EDE), Algorithm (ALG) and Artificial Intelligence (AI) are 12, 6, 9 respectively. Now we can use the Naïve Bayes classifier for probability calculation. The calculation of first term is based on the fraction of each target class in the training data. Prior probability for EDE, ALG and AI are 0.444, 0.333 and 0.222 respectively. The second term is calculated according to the equation (1). The probability values of word set are listed in Table 2.

**Table 2:** Word set with probability value

| Maximum Length Set | Probability | | |
|---|---|---|---|
| | EDE | ALG | AI |
| neural, network | 0.024 | 0.024 | 0.146 |
| gray, code | 0.043 | 0.217 | 0.043 |
| set, length | 0.043 | 0.130 | 0.043 |
| expectation, teacher, found, significant | 0.176 | 0.058 | 0.058 |
| education, level, test, significant, difference | 0.176 | 0.058 | 0.058 |
| significant, study, result | 0.235 | 0.058 | 0.058 |
| total, score, result | 0.176 | 0.058 | 0.058 |
| student, achievement, data, revealed | 0.176 | 0.058 | 0.058 |
| ratio, hole | 0.117 | 0.058 | 0.058 |
| gain, dependencies | 0.024 | 0.024 | 0.073 |
| Lossless, compression | 0.024 | 0.024 | 0.073 |

### 4.4 Proposed Algorithm

n = number of class, m = number of associated sets

1. For each class i = 1 to n do
2. Set pval = 0, nval = 0, p = 0, n = 0
3. For each set s = 1 to m do
4. If the probability of the class (i) for the set (s) is maximum then increment pval else increment nval
5. If 50% of the associated set s is matched with the keywords set do step 6 else do step7
6. If maximum probability matches the class i then increment p
7. If maximum probability does not match the class i increment n
8. If (s<=m) go to step 3
9. Calculate the percentage of matching in positive sets for the class i
10. Calculate the percentage of not matching in negative sets for the class i
11. Calculate the total probability as the summation of the results obtained from step 9 and 10 and also the prior probability of the class i in set s
12. If (i<=n) go to step 1
13. Set the class having the maximum probability value as the result

### 4.5 Flow Chart of the Proposed Algorithm

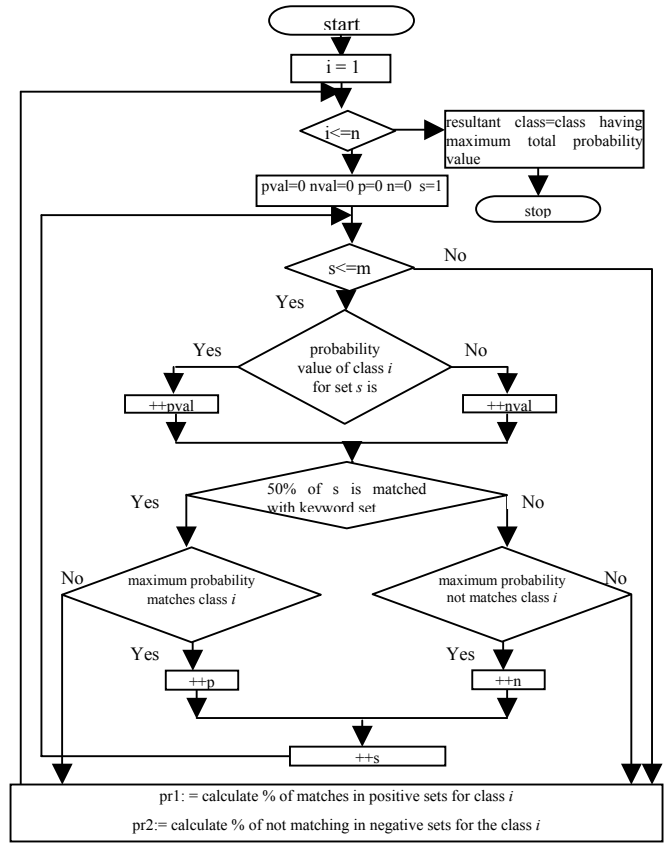

**Figure 1.** Flow chart of the proposed algorithm

## 5. COMPARATIVE STUDY

### 5.1 Association Rule and Naïve Bayes Classifier

The following results are found using the same data sets for both Association Rule with Naive Bayes Classifier and proposed method. The result shows that proposed approach works well using only 50% training data.

**Table 3:** Comparison of proposed method with text classifier using association rule and Naïve Bayes classifier

| % of Training Data | % of Accuracy | |
|---|---|---|
| | Association Rule with Naïve Byes Classifier | Proposed Method |
| 10 | 36 | 13 |
| 20 | 40 | 76 |
| 30 | 63 | 85 |
| 40 | 63 | 71 |
| 50 | 31 | 78 |

Accuracy Vs Training Data

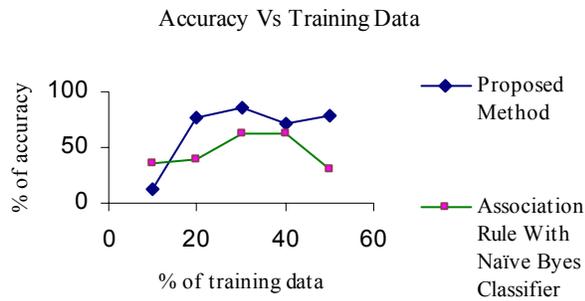

**Figure 2.** Accuracy Vs Training Curve

### 5.2 Association Rule Based Decision Tree

In text categorization using association rule based decision tree [11] 76 % data were used to train and the observed error was 13%. On the other hand using only 50% data as training, the proposed algorithm is able to classify text with 78% accuracy rate. The major problem of decision tree based classifier [8] [11] is that, this system is totally failed to categorize a class. Our proposed technique shows better performance even with 3 times larger data sets.

### 5.3 Genetic Algorithm

Researchers showed 68% accuracy using the concept of genetic algorithm with 31% test data [6] while our technique is better both in accuracy and % of test data. Moreover it required processing for each class during training. But our proposed algorithm does not require such process during training phase and hence reduces time.

**Table 4:** Comparison of proposed method with text classifier using decision tree and genetic algorithm

| Technique | (%) Training Data | (%) Accuracy |
|---|---|---|
| Association Rule Based Decision Tree | 76 | 87 |
| Genetic Algorithm | 69 | 68 |
| Proposed Algorithm | 50 | 78 |

## 5. CONCLUSION

This paper presented an efficient technique for text classification. The existing techniques require more data for training as well as the computational time of these techniques is also large. In contrast to the existing one, the proposed algorithm requires less training data and less computational time. In spite of the randomly chosen training set we achieved 78% accuracy for 50% training data. Though 85% accuracy was observed in 30% training data, a class could not be classified, so we dropped this position and increased training data set for more acceptable result. Though the experimental results are quite encouraging, it would be better if we work with larger data sets with more classes.